# Study on the frequency tuning of half-wave resonator at IMP*

HE Shou-Bo[1;2;1)] HE Yuan[1] ZHANG Sheng-Hu[1] ZHANG Cong[1] YUE Wei-Ming[1] WANG Zhi-Jun[1]
ZHAO Hong-Wei[1] LIU Lu-Bei[1] WANG Feng-Feng[1]

Institute of Modern Physics, Chinese Academy of Sciences, Lanzhou 730000, China
University of Chinese Academy of Sciences, Beijing 100049, China

**Abstract** A 162.5 MHz superconducting half-wave resonator (HWR) with geometry beta of 0.09 is being developed for Injector II of China Accelerator Driven Sub-critical System (CADS) Project at the Institute of Modern Physics (IMP). The HWR section composed of 16 HWR cavities will accelerate the proton beam from 2.1 MeV to 10 MeV. The RF and mechanical coupled analysis are essential in geometry design in order to predict the deformation of the cavity walls and the frequency shift caused by the deformation. In this paper, the detuning caused by both bath helium pressure and Lorentz force is analysed and a tuning system has been investigated and designed to compensate the detuning by deforming the cavity along the beam axis. The simulations performed with ANSYS code show that the tuning system can adjust and compensate the frequency drift due to external vibrations and helium pressure fluctuation during operation.
**Key words** CADS, HWR, vibration, tuning, ANSYS
**PACS** 29 20.Ej, 28.65.+a

## 1  Introduction

A high current superconducting proton linac has been designed for the China Accelerator Driven Sub-critical System (CADS) undertaken by the Chinese Academy of Sciences since 2011 and two different injectors are being fabricated by the Institute of High Energy Physics (IHEP) and the Institute of Modern Physics (IMP)[1]. The Injector II of CADS is being designed and built by IMP. As shown in Fig. 1, Injector II consists of the ECR ion source, the low energy beam transport line (LEBT), the radio frequency quadruple (RFQ), the medium energy beam transport line (MEBT) and the superconducting accelerating section. For the superconducting accelerating section, 16 superconducting coaxial half-wave resonator (HWR) cavities are elaborately designed and arranged to accelerate the proton beam from 2.1 MeV to 10 MeV.

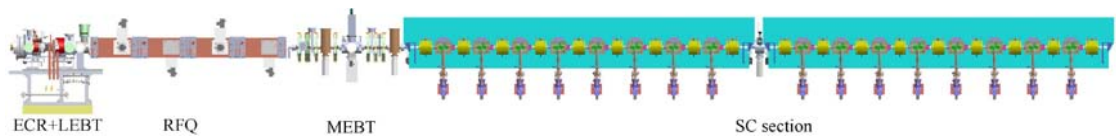

Figure 1. The schematic layout of Injector II for CADS.

Superconducting (SC) coaxial HWR was first proposed and prototyped in the early 1990s [2]. Recently, superconducting HWR structures are widely adopted in several proposed or proved high-intensity light ion linac projects [3,4,5]. Compared with the widely used SC quarter-wave resonator (QWR)[6], which has a more compact structure, one of the main advantages of HWR cavity is the geometry symmetry in vertical direction so that there is no vertical beam steering effect, and is suitable for accelerating high intensity light ion beam. As to spoke-type resonators, the HWR cavity is easier in fabrication and more stable, so it can obtain a higher real-estate gradient, and all these make it a possible alternative choice for filling the gap between RFQ and superconducting elliptical cavities[7].

*supported by Strategic Priority Research Program of CAS (XDA0302)and National Natural Science Foundation of China (91026004)
1)heshoubo@gmail.com

However, as the load quality factor of SC cavity usually is extremely higher than that of normal conducting cavity, SC HWR cavities are highly sensitive to mechanical deformations due to the narrow bandwidth.

The designed RF parameters of the HWR cavity [8] for Injector II are shown in Table 1. Due to the narrow bandwidth, a frequency tuning system is proposed. In Section III of this paper, general considerations on the design of this tuning system, including helium pressure influence and Lorentz force detuning effect, are discussed based on the calculating results with the ANSYS Code [9]. The detuning range and the upper limitation on accelerating gradient of this cavity are obtained. In the fourth section, the mechanical design based on the results of Section III is presented, and conclusions and suggestions for future research are given in the last section.

Table 1. The main RF parameters of the HWR cavity.

| Parameter | Value |
|---|---|
| Frequency/MHz | 162.5 |
| $\beta$ ( optimization) | 0.1 |
| $U_{acc}$/MV($B_{peak}$=50mT) | 0.78 |
| $B_{peak}/E_{acc}$[1](mT/MV/m)) | 10.92 |
| $E_{peak}/E_{acc}$ | 5.34 |
| $R/Q_0$ /$\Omega$ | 148 |
| $G=R_s*Q_0$/$\Omega$ | 28.5 |
| $Q_0$(K=4.4) | 1.4E9 |

1) The effective distance is the length between iris ($2\beta\lambda/3$).

## 2  Basic model setup

The structure of the superconducting HWR cavity with frequency of 162.5MHz which will be discussed is shown in Figure 2. Four rinsing ports are designed to guarantee the BCP processing and easy access to the inner conductor surfaces during the high pressure rinsing (HPR).

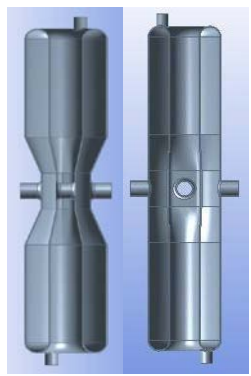

Figure 2. The structure of the HWR cavity.

The properties of niobium used in the coupled analysis in the following are listed in Table 2. It is worth pointing out that the thickness of niobium for the HWR cavity has changed to 2.8 mm because the BCP technique etches about 200 μm of the cavity surface.

Table 2. The properties of niobium for SC HWR.

| | |
|---|---|
| Density/(g/cm$^3$) | 8.6 |
| Tensile strength /MPa | 400 |
| Poisson's ratio | 0.38 |
| Young's modulus/GPa | 105 |
| Thickness/mm | 2.8 |

The thickness of cavity wall has an influence on the level of frequency detuning. Due to the asymmetry of HWR cavity, all of the simulation and analysis are based on the results of the whole cavity calculation.

## 3  General considerations on the tuning system design

During the mechanical analysis, the variation of eigenmode frequency was calculated. The relationship between changes in RF frequency and mechanical deformations is given by the Slater Perturbation Theorem [10].

$$\frac{\Delta f}{f_0} = \frac{\int_{\Delta v} (\mu_0 |\vec{H}_0(\vec{x})|^2 - \varepsilon_0 |\vec{E}_0(\vec{x})|^2) d^3 x}{4 \cdot U_0} \quad (1)$$

From Equation (1) we can see that the deformation in the electric field region and the deformation in the magnetic field region will give an opposite contribution to the frequency shift.

The dominant source of mechanical deformations comes from the vibrations due to various external sources. Generally sources of frequency detuning for SC cavity include (i) tolerances in the cavity parts; (ii) variation in welding shrinkage; (iii) variation in the frequency due to the attachment of the helium vessel; (iv) variation in the frequency produced by etching; (v) variation in the frequency during the cooling down; (vi) variation in the frequency shift by beam loading; (vii) variation in the liquid helium bath pressure; and (viii) Lorentz force detuning. All these detuning effects should be taken into account during the frequency tuning before horizontal test. Since most of the sources can be avoided by better machining and fabrication, in this paper we will mainly discuss about the liquid helium pressure sensitivity and Lorentz force detuning effect.

### 3.1  Pressure sensitivity

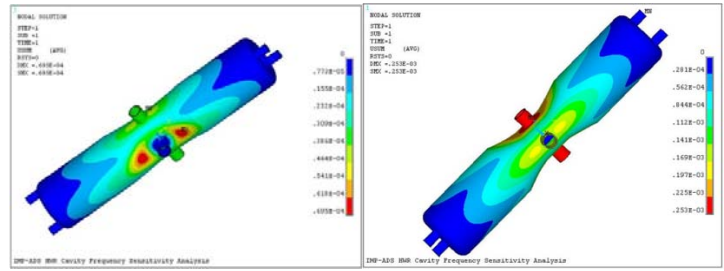

Figure 3. The deformation results of HWR with beam pipes fixed (left) and beam pipes free (right).

One source of vibrations, operating at 4.2 K, is that the fluctuations in the local helium bath pressure from the evaporation caused by the RF losses [11]. This pressure difference deforms the cavity geometry and shifts the cavity frequency. As the cavity stiffening condition is hard to determine, it is necessary to perform this research under different boundary conditions at the two cavity beam ports, named fully fixed condition and completely free condition, to estimate the pressure sensitivity. During the calculation, the tetrahedron mesh with a total grid number of 1186751 is used, which can guarantee the accuracy of the analysed results. The same meshed model was used for all types of simulations to evaluate the frequency shift at the same scale. The simulated cavity deformations under one atmosphere pressure are shown in Fig. 3, which illustrate that the main frequency shift is caused by the beam port displacement. The maximum deformation is located in the middle of the outer conductor, which is near the high electric field region. According to Eq.1, the frequency shift will decrease. The pressure sensitivities for the HWR with different boundary conditions are listed in Table 3.

Table 3. The simulation results for HWR cavity under one atmosphere load.

| Boundary | Displacement/mm | Stress/ksi | $\Delta f$ /KHz | df/dp/(Hz/torr) |
|----------|-----------------|------------|-----------------|-----------------|
| fix      | 0.0695          | 5.786      | 17.892          | 23.54           |
| free     | 0.252           | 19.066     | 110.773         | 145.6           |

During the operation, the boundary is dynamically balanced between the fixed and the free conditions. In this case, the pressure sensitivity coefficient, df/dP, is between 23.54 Hz/torr and 145.6 Hz/torr. Because the fluctuation range of helium bath pressure in cryogenic plant is about ±3 mbar, according to the experience of IMP, the frequency drift will vary in the range of 53.09Hz to 329.93Hz, which is in the same order of magnitude of HWR cavity bandwidth 235Hz. For the HWR tuner design, the tuning range and tuning resolution should be taken into account. In order to improve the pressure sensitivity coefficient df/dP, more efforts will be focused on adding some stiffening ribs on the cavity in the future.

### 3.2 Lorentz force detuning (LFD) effect

Lorentz force is the result of the interaction between the electromagnetic field in a cavity and its RF wall current. In this circumstance, the local electromagnetic field will be redistributed. According to Eq. (1), this effect can be described by the Lorentz force detuning coefficient $K_L$:

$$K_L = \Delta f / E_{acc}^2 \qquad (2)$$

From this formula, the frequency shift $\Delta f$ is proportional to the square of accelerating gradient. The resonant frequency will correspondingly decrease with the increase of accelerating gradient. The numerical 3D analysis for the LFD effect has been done and an initial simulation for the cavity with fully fixed beam pipes is shown in Figure 4. The maximum deformation is located near the high electric field region.

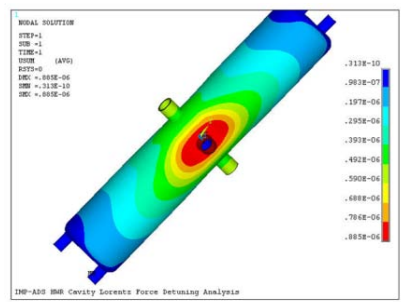

Figure 4. The deformation of the cavity wall caused by LFD effect.

The relationship between frequency shift and accelerating gradient $E_{acc}$ is plotted in Fig. 5. According to the final fitting curve between $\Delta f$ and $E_{acc}$ in Fig. 5, the Lorentz force detuning coefficient $K_L$ is -4.657 Hz/(MV/m)$^2$. From the point of view of energy content, the frequency detuning is still acceptable if the accelerating gradient is less than 5 MV/m. When the accelerating gradient is beyond 5 MV/m and the frequency shift is larger than -116 Hz, which is about half the frequency bandwidth, it means that stiffener is required.

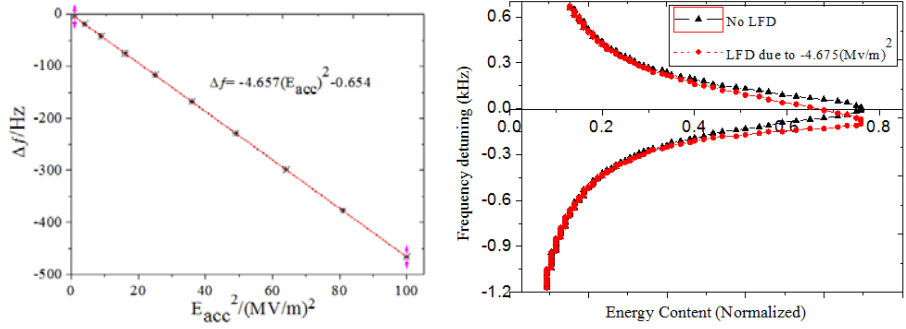

Figure 5. The frequency shift vs. accelerating gradient (left); the relationship between LFD (Eacc=5MV/m) and energy content (right).

## 4  Tuning System

Generally speaking, we can enhance the cavity rigidity by increasing the wall thickness and reducing the frequency shift. However, it will increase the cost and inefficient for cooling with liquid helium. Besides this method, two approaches are usually adopted, one is extra power and the other is mechanical tuner. We will discuss about them separately in the following section.

The relationship between generator power and external Q value with different acceleration gradients is illustrated in Fig. 6. Extra power is needed to meet the requirement of the designed accelerating gradient. Obviously, extra power for the compensation can be well reduced under proper pre-detuning.

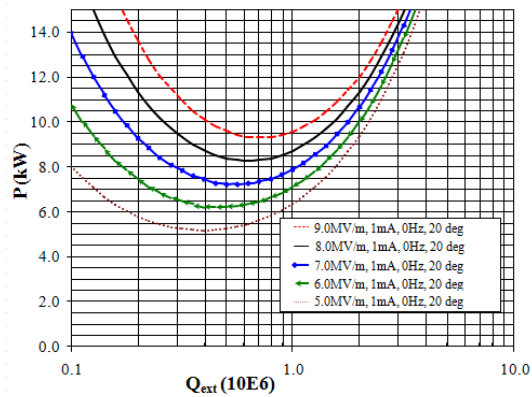

Figure 6. The relationship between generator power and external Q value (different accelerate gradients).

Basically, there are two methods for mechanical tuner. The first one is to mechanically deform the cavity. The other technique is to insert a probe into the magnetic or electric field of SC cavity. The tuning method adopted by IMP is to change the length of the cells by mechanical adjustment of the overall length of the cavity, which introduces no new HOMs [12]. Figure 7 shows the structure of the mechanical tuner for HWR cavity.

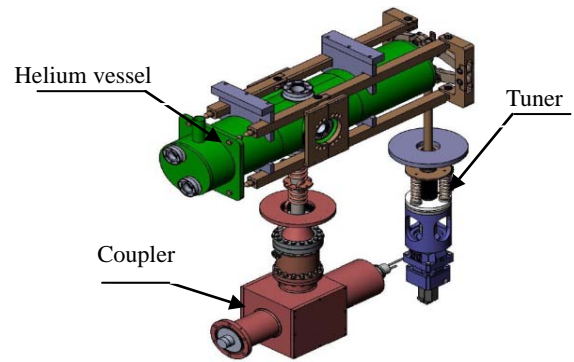

Figure 7. The HWR mechanical tuner structure.

The scissor jack of the tuner is mechanically mounted on the helium vessel and it will provide slow tuning. The tuner mechanism consists of four levers and twelve anti-backlash pivots, which is driven by a step motor. The main parameters obtained during the design of the tuner are listed in Table 4. Considering the fabrication tolerances and microphonics detuning of the cavity, the tuner is designed to have the ability to retune the cavity over a range of 360 kHz centred on the nominal operation frequency of 162.5 MHz. The tuning range and resolution can accommodate differences between the designed and the operated frequencies.

Table 4. The designed parameters of the tuner system.

| Parameter | Value |
| --- | --- |
| Tuning sensitivity/(KHz/mm) | 180 |
| Tuning force/(KN/mm) | 2.2 |
| Tuning range/KHz | 360 |
| Tuning resolution/Hz | 9 |
| Tuning resolution step/nm | 50 |
| Fine Tuning Range/Hz | 180 |
| Backlash Tolerance/Hz | <2 |

## 5  Summary

The mechanical and RF field coupled analysis of the HWR cavity has been performed at IMP. Both the helium pressure influence and Lorentz force detuning effect, which are important and crucial reasons in the frequency detuning, are studied in this paper. A reliable tuner has been designed that meets the requirements for compensating the frequency drift due to various external sources. A prototype tuning system for experiment has been fabricated recently and will be improved in the following measurement. Stiffening ribs are supposed to be added to the cavity walls in the future in order to tune the cavity more effectively.

*One of the authors, He Shou-Bo, would like to thank Prof. Haipeng Wang and Dr. Gary Cheng of JLAB for the idea of simulations and helpful discussions.*